# Bipolar Transistor Based on Graphane

**B Gharekhanlou, S B Tousaki, S Khorasani**

School of Electrical Engineering, Sharif University of Technology,
PO Box 11365-9363, Tehran, Iran.

E-mail: khorasani@sharif.edu

**Abstract.** Graphane is a semiconductor with an energy gap, obtained from hydrogenation of the two-dimensional grapheme sheet. Together with the two-dimensional geometry, unique transport features of graphene, and possibility of doping graphane, p and n regions can be defined so that p-n junctions become feasible with small reverse currents. Our recent analysis has shown that an ideal I-V characteristic for this type of junctions may be expected. Here, we predict the behavior of bipolar juncrion transistors based on graphane. Profiles of carriers and intrinsic parameters of the graphane transistor are calculated and discussed.

## 1. Introduction

Graphane is an extended two-dimensional (2D) covalently bonded hydrocarbon. It is a fully saturated hydrocarbon derived from a single graphene sheet with formula $C_6H_6$. All of the carbon atoms are in $sp^3$ hybridization forming a hexagonal network and the hydrogen atoms are bonded to carbon on both sides of the plane in an alternating manner. It has a relatively large direct energy gap of about 3.6eV [1-6]. The numerical estimate for the intrinsic carrier density of graphane turns out to be too low for practical applications [7]. One of the simplest ways to reduce the energy gap is to bring up hydrogen deficiency in graphane structure [7]. So we use $C_6H_3$ Structure. In this paper, we analyze a 2D bipolar transistor based on doping of a modified graphane sheet. The graphane structure is supposed to encompass hydrogen deficiency, with a typical stoichiometric composition of $C_6H_{6-x}$. We have used the tight-binding method to show how hydrogen deficiency may be exploited to reduce the band gap of graphane from 3.6eV down to about 1.1eV [7]. This increases the intrinsic carrier density and thereby the conductivity of doped graphane. Analysis of the junctions with the aid of Shockley's law is done to obtain the profiles and behaviour of the corresponding bipolar transistor.

## 2. Analysis

Necessary parameters are calculated at 300K [7]. The density of states for conduction and valence bands are respectively $N_c=2.1\times10^{17}m^{-2}$ and $N_v=2.6\times10^{17}m^{-2}$, with the intrinsic carrier density of $n_i=4.5\times10^7m^{-2}$. Now, we proceed with a $p^+np$ bipolar junction transistor, with the dopant-concentrations in the emitter, base and collector respectively given by $N_E=2.0\times10^{15}m^{-2}$, $N_B=10^{14}m^{-2}$, and $N_C=10^{13}m^{-2}$. One can immediately use the relation $V_b=V_T\ln(N_AN_D/n_i^2)$ with $V_T=K_BT/q$, to readily get $V_{jBE}=0.8340V$ and $V_{jBE}=0.6968V$ for emitter-base and base-collector junctions, respectively.

As we deal with 2D junctions, the analysis the electric field is different to that of bulk step junctions [7]. For this purpose, we need to divide the charge sheets into infinitely many thin wire line charges, with constant charge densities $\lambda_i$ [7]. The partial electric field due to each line charge is $E_i(r)=\lambda_i/2\pi\varepsilon\, r_i$, where $r_i$ is the distance from the charge wire. This model is illustrated in figure 1. Integrating from the electric feild gives the potential distribution. Calculated total electric field over the surface of the transistor is demonstrated in figure 2. We also find the width of the deplation regions in the emitter, base and collector to be in the order of micrometer.

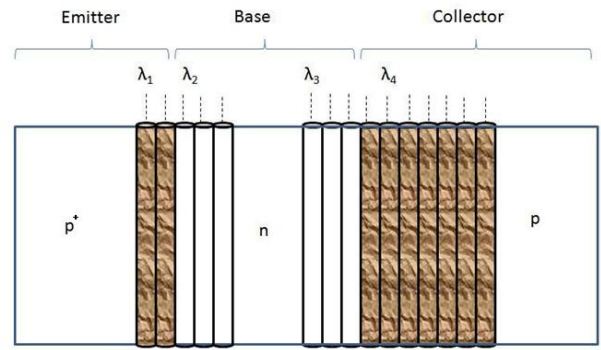

**Figure 1**. Schematic diagram of a pnp bipolar transistor.

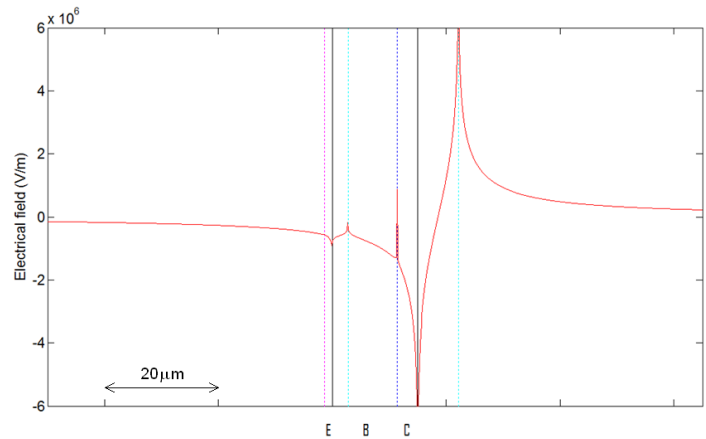

**Figure 2**. The electric field distribution in the graphane bipolar transistor.

### 3. Current-voltage characteristics
Figure 3 represents a schematic of a pnp transistor biased in the common-base configuration. It also indicates all the relevant current components of the transistor in the circuit.

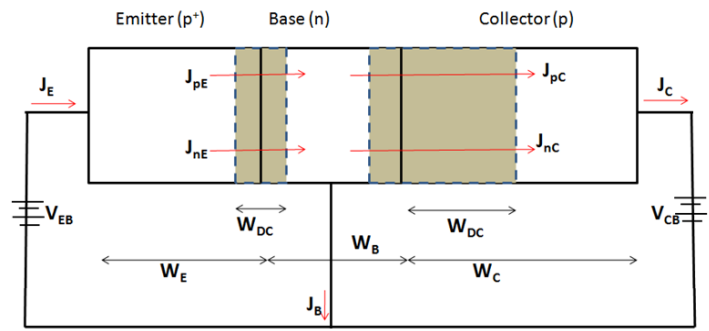

**Figure 3**. Current components of the graphane pnp transistor.

In the neutral base region, the injected minority-carriers distribution (holes) is governed by the continuity equation, with the boundary conditions at the two edges of neutral region

$$-\frac{p_n - p_{n0}}{\tau_p} - \mu_p E \frac{dp_n}{dx} + D_p \frac{d^2 p_n}{dx^2} = 0 \tag{1.1}$$

$$p_{nB}(x = x_{nE0}) = n_{nB0} \exp\left(\frac{qV_{EB}}{K_B T}\right) \tag{1.2}$$

$$p_{nB}(x = w - x_{nC0}) = n_{nB0} \exp\left(\frac{qV_{CB}}{K_B T}\right) \tag{1.3}$$

The general solution to the above equation is given by

$$p_n(x) = p_{n0} \exp\left(\frac{x + x_{nC}}{L_{p1}}\right) \left[-\exp\left(\frac{w}{L_{p2}}\right) + \exp\left(\frac{qV_{EB}}{K_B T} + \frac{w}{L_{p2}}\right) + \exp\left(\frac{x_{nE} + x_{nC}}{L_{p2}}\right) - \exp\left(\frac{qV_{CB}}{K_B T} + \frac{x_{nC} + x_{nE}}{L_{p2}}\right)\right] \times$$

$$\left[\exp\left(\frac{w}{L_{p2}} + \frac{x_{ne} + x_{nC}}{L_{p1}}\right) - \exp\left(\frac{w}{L_{p1}} + \frac{x_{ne} + x_{nC}}{L_{p2}}\right)\right]^{-1}$$

$$+ p_{n0} \exp\left(\frac{x}{L_{p2}}\right) \left\{-\exp\left(\frac{x_{ne}}{L_{p1}}\right)\left[1 - \exp\left(\frac{qV_{CB}}{K_B T}\right)\right] + \exp\left(\frac{w - x_{nC}}{L_{p2}}\right)\left[1 - \exp\left(\frac{qV_{EB}}{K_B T}\right)\right]\right\} \times \tag{2}$$

$$\left[\exp\left(\frac{w - x_{nC}}{L_{p2}} + \frac{x_{nE}}{L_{p1}}\right) - \exp\left(\frac{w - x_{nC}}{L_{p1}} + \frac{x_{nE}}{L_{p2}}\right)\right]^{-1} + p_{n0}$$

The hole density in the base region is used for calculation of hole current due to the collector and emitter depletion regions. The electron currents at the emitter edge $J_{nE}$ and the collector edge $J_{nC}$ are given by

$$J_{nE} = -qD_n \frac{dn_{pe}}{dx}\bigg|_{x=-x_{pe0}} = \frac{qD_n n_{pe0}}{L_{n2}(x)}\left[\exp\left(\frac{qV_{eb}}{K_B T}\right) - 1\right]\exp\left[\frac{x + x_{pe0}}{L_{n2}}\right]\bigg|_{x=-x_{pe0}}$$

$$= \frac{qD_n n_{pe0}}{L_{n2}(x = -x_{pe0})}\left[\exp\left(\frac{qV_{eb}}{K_B T}\right) - 1\right] \tag{3.1}$$

$$L_{n2}(x) = \left\{\frac{\mu_n}{2D_n} E(x) + \sqrt{\left[\frac{\mu_n}{2D_n} E(x)\right]^2 + \frac{1}{\tau_n D_n}}\right\}^{-1} \tag{3.2}$$

$$J_{nC} = -qD_n \frac{dn_{pC}}{dx}\bigg|_{x=-x_{pC0}} = \frac{qD_n n_{pC0}}{L_{n2}(x)}\left[\exp\left(\frac{qV_{CB}}{K_B T}\right) - 1\right]\exp\left[\frac{x - (w + x_{pc0})}{L_{n1}}\right]\bigg|_{x=-x_{pC0}}$$

$$= \frac{qD_n n_{pC0}}{L_{n1}(x = -x_{pC0})}\left[\exp\left(\frac{qV_{CB}}{K_B T}\right) - 1\right] \tag{3.3}$$

$$L_{n1}(x) = \left\{\frac{\mu_n}{2D_n} E(x) - \sqrt{\left[\frac{\mu_n}{2D_n} E(x)\right]^2 + \frac{1}{\tau_n D_n}}\right\}^{-1} \tag{3.4}$$

The emitter and collector currents can be then obtain from

$$J_E = J_{nE} + J_{pE} \tag{4.1}$$

$$J_C = J_{nC} + J_{pC} \tag{4.2}$$

Here $J_{pE}$ and $J_{pC}$ are minority-carriers current densities in the base region, flowing to emitter and collector regions, respectively. Figure 4 shows a typical collector current characteristics, which confirms the exponential current-voltage dependence of the graphane transistor. Here, the collector-base junction is biased at $V_{CB} = -1$V. Figure 5 shows a representative set of output characteristics for the common base configuration.

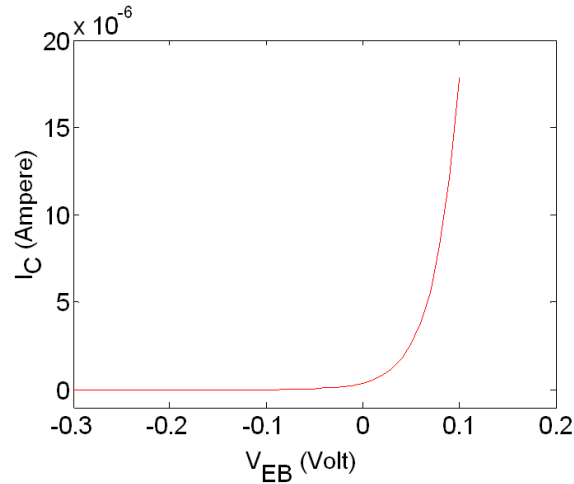

**Figure 4**. Collector current characteristics versus emitter-base voltage.

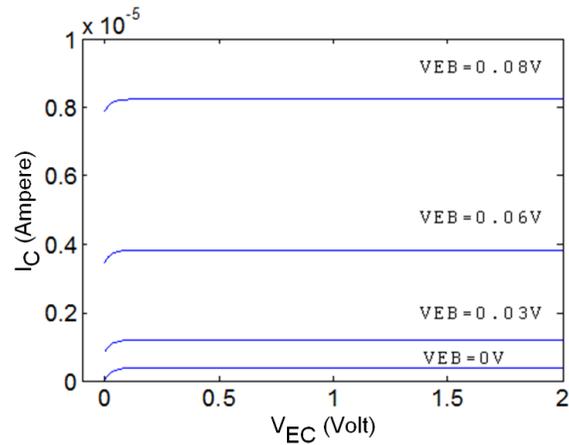

**Figure 5**. Output characteristics for common base configuration.

## 4. Conclusions
We have introduced a 2-D p$^+$np bipolar transistor based on graphane with hydrogen deficiency to reduce the band gap effectively. Using a basic analysis, we have shown that, within the approximation of the Shockley's law of junctions, an exponential ideal I–V characteristic is expectable. Furthermore, we have calculated the surface distribution of electric field in the vicinity of the junction. Curvature of collector current characteristics shows good agreement with an ideal bipolar transistor.